\documentstyle[12pt,aasms4]{article}
\begin{document}

\begin{center}
\title{The Nature of the Secondary Star in the Black Hole X-Ray Transient V616 Mon (=A0620-00)}

\author{Thomas E. Harrison$^{\rm 1,2,3}$}

\affil{Department of Astronomy New Mexico State University, Box 30001, MSC 4500, Las Cruces, NM 88003-8001}

\authoremail{ tharriso@nmsu.edu}

\author{Steve B. Howell}

\affil{WIYN Observatory and National Optical Astronomy Observatories, 950 North Cherry Avenue, Tucson, AZ 85726}

\authoremail{howell@noao.edu}

\author{Paula Szkody$^{\rm 1}$}

\affil{Department of Astronomy, University of Washington, Box 351580, Seattle, WA 98195}

and

\author{France A. Cordova}

\affil{Institute of Geophysics and Planetary Physics, Department of Physics,
University of California, Riverside, CA 92521}

\end{center}

\noindent
$^{\rm 1}$Visiting observers, W. M. Keck Observatory which is operated as a 
scientific partnership among the California Institute of Technology, the 
University of
California and the National Aeronautics and Space Administration.\\
$^{\rm 2}$Visiting Astronomer, Kitt Peak National Observatory, National
Optical Astronomy Observatory, which is operated by the Association of 
Universities for Research in Astronomy, Inc., under cooperative agreement with
the National Science Foundation.\\
$^{\rm 3}$Visiting Astronomer at the Infrared Telescope Facility, which is
operated by the University of Hawaii under contract from the National
Aeronautics and Space Administration.\\


\abstract{ We have used NIRSPEC on Keck II to obtain $K$-band spectroscopy of
the low mass X-ray binary V616 Mon (= A0620$-$00). V616 Mon is the 
proto-typical soft x-ray transient containing a black hole primary. As such
it is important to constrain the masses of the binary components. The modeling of
the infrared observations of ellipsoidal variations in this system lead to
a derived mass of 11.0 M$_{\sun}$ for the black hole. The validity of this
derivation has been called into question due to the possiblity that the
secondary star's spectral energy distribution is contaminated by accretion
disk emission (acting to dilute the variations).  Our new $K$-band 
spectrum of V616 Mon reveals a late-type K dwarf secondary 
star, but one that has very weak $^{\rm 12}$CO absorption features. Comparison of
V616 Mon with SS Cyg leads us to estimate that the accretion disk supplies 
only a small amount of $K$-band flux, and the ellipsoidal variations
are not seriously contaminated. If true, the derived orbital inclination of V616
Mon is not greatly altered, and the mass of the black hole remains large.
A preliminary stellar atmosphere model for the $K$-band spectrum of V616 Mon 
reveals that the carbon abundance is approximately 50\% of the solar value. We 
conclude that the secondary star in V616 Mon has either suffered serious
contamination from the accretion of supernova ejecta that created the black 
hole primary, or it is the stripped remains of a formerly more massive 
secondary star, one in which the CNO cycle had been active.}

\section{Introduction}

V616 Mon (= A0620$-$00) is the proto-type for a small family of binaries
known as Black Hole X-ray Transients (BHXTs) where the primary is a 
black hole, and the secondary a relatively low mass star. A sudden increase in the mass transfer rate due to an accretion 
disk instability has been used to explain the outbursts of these systems 
(Cannizzo 1993, Menou 2002). After a number of months, the BHXTs return to
relative quiescence and are very faint at X-ray and optical wavelengths.
During this time, both their broadband spectral energy distributions
(SEDs) and moderate resolution spectra are dominated by a late-type secondary star (c.f., Gelino
et al. 2001, Gonz$\acute{a}$lez Hern$\acute{a}$dez et al. 2004). Because
the secondary star is clearly visible, radial
velocity observations are possible, allowing the derivation of its
orbital and rotational motion. When combined with the 
orbital period, and an estimate for the orbital inclination, it is possible to 
put useful limits on the masses of both components in the system. When this
is done for V616 Mon, the mass of the black hole primary is 11 M$_{\sun}$.
As such, V616 Mon (P$_{\rm orb}$ = 7.75 hr) has the largest
mass black hole for any X-ray binary whose orbital period is 
P$_{\rm orb}$ $\leq$
24 hr (Lee et al. 2002). As discussed by Kalogera (1999), it is difficult
to produce a binary with a black hole primary that has such an extreme mass ratio.

In the case of V616 Mon, the leading uncertainty in the derivation of the black hole mass is the
orbital inclination angle. Measurement of the orbital inclination in any close 
binary system depends
on some assumptions about the nature of the stellar components. But with
estimates for the gravity and effective temperature of the stars in such
systems, it is possible to use light curve modeling software (e.g., that of 
Wilson-Divinney 1998 = WD98; see Kallrath et al. 1998) to fit the orbital 
inclination if it 
exhibits eclipses, or if one of the components is distorted enough to produce 
ellipsoidal variations. Since the outbursts of BHXTs are believed to be 
due to an accretion disk event, the secondary star must be acting as the
donor in such systems, and can safely be assumed to fill its Roche lobe. 
Therefore BHXTs will exhibit ellipsoidal variations if the orbit is inclined to 
our line of sight, and the secondary star emits a significant fraction of
the systemic luminosity. As shown by Shahbaz et al. (1994) and Gelino et al. 
2001, V616 Mon exhibits such ellipsoidal variations in the near-IR. Using 
published parameters for the system derived from earlier radial velocity 
studies, Gelino et al. concluded that the system has an orbital 
inclination of 41$^{\circ}$ $\pm$ 3.  In that work, Gelino et 
al. found that the existing $BVRIJHK$ photometry of V616 Mon was consistent 
with a K4V secondary star, and assumed that the $JHK$ ellipsoidal variations
were relatively uncontaminated by emission from the accretion disk and/or
hot spot. Hynes et al. (2005) have recently cast doubt on the efficacy of
such a model by exploring the possibility that the accretion disk
severely contaminates the SED, and might even mimic a K-type secondary star.
They justify this exercise by reference to an existing $K$-band spectrum
of this source by Shahbaz et al. (1999) that seemed to lack obvious 
CO absorption features. Below, we present a new, higher resolution and higher
S/N infrared spectrum of V616 Mon where we find a late K-type 
stellar spectrum, one in which the $^{\rm 12}$CO features are 
anonymously weak. 
We then present preliminary stellar atmosphere models for this object,
from which we derive estimates for the $^{\rm 12}$C abundance, and discuss 
these results in the context of existing BHXT formation scenarios.

\section{Observations}

V616 Mon was observed on 2005 February 17 using NIRSPEC\footnote{For more on NIRSPEC go to www2.keck.hawaii.edu/inst/nirspec/nirspec.html} on
Keck II in photometric conditions. We used NIRSPEC in low resolution mode 
with a 0.38" slit.  The grating tilt was set so as to cover the wavelength 
region 2.04 $\mu$m $\leq$ $\lambda$ $\leq$ 2.46 $\mu$m, with a dispersion of 
4.27 \AA ~pixel$^{\rm -1}$. We employed the two-nod script, and obtained
twelve individual exposures with four minute integration times. To correct for telluric absorption, we 
observed bright A0V stars located close to the program object so as to 
minimize their relative differences in airmass. We used standard IRAF
spectroscopic data reduction routines to produce the final spectra. In the $K$-band, the spectra of A0V stars are nearly featureless, except
for the prominent H I Brackett $\gamma$ absorption line at 2.16 $\mu$m. To 
remove the telluric features we divide by the spectrum of this A star. This
process artificially enhances the H I emission. To allow for an examination of 
the profile of the H I Brackett $\gamma$ emission line, we have patched-over 
the intrinsic H I absorption feature in the A0V star before division into
our program spectra.
 
Before we can coadd the twelve spectra, we must correct for the radial
velocity motion of the secondary star. Gelino et al. (2001) found that the 
observed ellipsoidal variations of V616 Mon were synchronized to the 
orbital ephemeris of Leibowitz et al. (1998). We have phased our spectra using 
that ephemeris. The final $K$-band spectrum of V616 Mon, obtained from the
median of the Doppler corrected spectra, is presented
in Fig. 1 where we compare it to IRTF spectra of two K dwarfs (from
Harrison et al. 2004). Note that the profiles of H I Brackett $\gamma$ line and
that of the He I line at 2.058 $\mu$m, are affected by the Doppler correction 
since
the origin of the line emission occurs interior to the orbit of the secondary
star.  All spectra have been smoothed to a dispersion of 20 \AA/pix, and the K 
dwarf spectra have been convolved to a rotation velocity of 83 km s$^{-1}$ to 
match the observed $v$sin$i$ of the secondary star in V616 Mon (Marsh, 
Robinson, \& Wood 1994). The S/N of our V616 Mon spectrum is lower than might 
be expected using NIRSPEC, but the seeing early in the night was very poor, FWHM 
$\sim$ 2". Fortunately it improved as the night wore on (see Harrison et al. 
2005).

\section{The Secondary Star of V616 Mon}

As is clear from Fig. 1, the spectrum of V616 Mon appears to be that
of a K-type dwarf, except for the presence of He I (at 2.058 $\mu$m) and
H I (at 2.166 $\mu$m) emission, and the weakness of its CO absorption features. 
The presence of the H I and He I emission lines indicates continued accretion
in the system. The equivalent widths of the H I (23.0 $\pm$ 1.2 \AA) and He I (12.0 +/- 1.7 \AA) lines remained constant over the small range in orbital phase 
covered by our observations (0.93 $\leq$ $\phi$ $\leq$ 1.00), except for a 
brief decline of 30\% in both lines at phase 0.98. However, our data only
cover 12\% of an orbital period for V616 Mon, so nothing can be said on how
repeatable this is, whether phase 0.975 is special, or if similar variations
occur at other times.  Shahbaz et al. (1999) 
measured an orbitally averaged equivalent width for H I of 14.6 $\pm$ 1.2 \AA, 
suggesting either increased accretion activity during our observations, or that 
the observed H I emission is higher near inferior conjunction of the secondary
star.

Previous estimates of the spectral type of the secondary star range from K3/4V 
(Haswell et al. 1993; Shahbaz et al. 1999) to K5/7V (Oke 1977). Gelino et al. 
(2001) assumed a spectral type of K4V. Careful visual comparison of our new 
spectrum of V616 Mon to our library of templates suggests a slightly later 
spectral type of K5V. This is more clearly shown by comparison of the strengths 
of Na I and Ca I equivalent widths listed in Table 1, as well as in the
trends of other atomic absorption features to have values closer to those of the
K5V template when compared to those of the K3V. It is important to note that 
Gonz$\acute{a}$lez Hern$\acute{a}$dez et al. (2004) concluded that iron
and aluminum were slightly enhanced (as were Li, Ti and Ni) in the secondary 
star of V616 Mon. The equivalent widths we measure for iron and aluminum are 
consistent with those findings. Due to the large rotational velocity of
V616 Mon, a number of the lines for which we have tabulated equivalent widths 
in Table I are blends, making it impossible to deconvolve which species is 
responsible for the measured line strength. 

The continuum of V616 Mon does appear to be flatter than either of the
templates, having a slope of -1.1 ($\times$ $\lambda$ in $\mu$m), versus -1.5 
for the templates. The 
template spectra obtained at the IRTF were reduced using the 
SPEXTOOL program that employs a stellar atmosphere model for telluric 
correction, while our Keck data were reduced assuming a blackbody spectrum for 
the A star. Thus, some of this difference in spectral slope could result from
these different reduction procedures. But spectra of a number of long
period CVs presented in Harrison et al. (2004) had flatter 
than expected spectral slopes if the secondary star was the sole source of
luminosity in the $K$-band. In each of those cases (e.g., SS Cyg), the CVs in 
question have detectable infrared emission from their accretion
disks (see below). Obviously, the presence of H I and He I emission in the 
spectrum of V616 Mon shows that accretion is continuing at a low level, and 
thus it is likely that the continuum slope of the infrared spectrum is affected.

The most prominent anomaly in 
the spectrum of V616 Mon is the weakness of the $^{\rm 12}$CO features, and the 
apparent enhancement of
the absorption from $^{\rm 13}$CO. While the $^{\rm 13}$CO$_{\rm (2,0)}$ 
feature at 2.345 $\mu$m is not well-defined, the next overtone, 
$^{\rm 13}$CO$_{\rm (3,1)}$ at 2.374 $\mu$m, is quite prominent. As shown
in Fig. 1., the $^{\rm 13}$CO features are insignificant in the spectra
of normal late-type dwarfs. Thus, we interpret its strength here as an 
enhancement over its normal abundance. We attempt to derive the abundances 
of both $^{\rm 12}$CO and $^{\rm 13}$CO in the next section.

\subsection{Derivation of the Isotopic Abundances of Carbon}

While the weakness of the $^{\rm 12}$CO features could be due to deficits
of either carbon or oxygen, an enhanced level of $^{\rm 13}$C would suggest
CNO cycle processing, the end result of which is enhanced levels of nitrogen
and $^{\rm 13}$C, reduced levels of $^{12}$C, with little change in the 
global oxygen abundance (Marks \& Sarna 1998). We have used the 
atmosphere program ``SPECTRUM'' by Gray \& Corbally (1994) to produce models 
for V616 Mon. SPECTRUM comes with an extensive line list (796782 lines!)
for modeling stellar spectra in the 1 to 4 $\mu$m region. Unfortunately,
trial spectra using this line list produced a large number of very strong
lines that are not seen in the template spectra, and appear to be due to 
scandium, vanadium, nickel, and other relatively rare atomic
species. Thus, we decided to construct our own line list from scratch
to model the CO lines, limiting ourselves to the region 2.28 to 2.42 $\mu$m. 
For this purpose we used the CO line list compiled by Goorvitch (1994). In 
addition, both Na I and Mg I have strong features in this region, so data 
for the Na I doublet at 2.34 $\mu$m and three lines from Mg I near 2.38 
$\mu$m were also included in our line list\footnote{The data for these
five transitions were obtained from the NIST Atomic Spectra Database at
physics.nist.gov/cgi-bin/AtData/lines\_form}. 

We found that in all of our initial trial model spectra calculated by SPECTRUM
using Kurucz atmospheres\footnote{Model atmospheres were obtained from 
Kurucz's website: kurucz.harvard.edu/grids.html} of the 
appropriate temperature, the spectral lines/features were 
weaker than they should be when compared to the template spectra. To fix this, we globally
elevated the oscillator strengths [the log(gf) values] of the CO, Na and Mg lines until we achieved
a good match between the model and template spectrum for the K5V. We then
constructed both hotter and cooler models using this line list and compared them
to our other template spectra to see if the spectral features evolved with 
temperature in the correct fashion. The result of this exercise is shown in
Fig. 2. The models fit our observed MK template spectra quite well considering
the limited line list, except for the match of the
$^{\rm 12}$CO$_{\rm (5,3)}$ feature at 2.383 $\mu$m to the spectra of the K5V 
and K7V. This CO bandhead is strongly affected by the Mg I absorption lines 
noted above.  We tested how sensitive the shape of this feature was
to the Mg I lines by increasing their oscillator strengths
and found that such models were better fits to the K5 and K7 spectra, but 
were worse for the hotter K2V and K3V templates (where the Mg I lines are more 
prominent relative to the $^{\rm 12}$CO$_{\rm (5,3)}$ feature). It is difficult to ascertain the origin of this discrepancy, given 
that both the hotter and cooler template spectra are well modeled. Perhaps the 
K5V and K7V templates (61 Cyg A and B) have slightly enhanced levels of 
magnesium.

With confidence that we could reproduce normal K dwarf spectra, we then ran
a series of K5V models (T$_{\rm eff}$ = 4500 K) with reduced levels of 
$^{\rm 12}$C and enhanced levels of $^{\rm 13}$C. We constructed models that
had carbon deficits that ranged from 0.9 to 0.1 solar. We attempted $\chi^{2}$ 
tests between
the model spectra and the observations, but such tests were inconclusive.
This is partly due to several important absorption features in this wavelength 
range missing from our models, but also from the low S/N of the data. The models 
that appear to $visually$ fit the observed spectrum of V616 Mon the best, shown 
in Fig. 3, are ones in which the $^{\rm 12}$C abundance is $\approx$ 50\% of 
solar. Unfortunately, our spectrum of V616 Mon is quite poor beyond 2.33 
$\mu$m, and it is difficult to be confident about the level of enhancement of 
$^{\rm 13}$C in our spectrum. A model where the isotopic ratio is 
$^{\rm 13}$C/$^{\rm 12}$C = 1, does a reasonable job of qualitatively fitting 
the observed features.

\section{Discussion}

Our new spectrum of V616 Mon resolves some of the earlier issues with the
nature of its secondary star. Clearly, the CO features in this object
are much weaker than they should be, while most of the atomic features are
at more normal strengths for a K5V. If the weakness is assumed to be
due to a deficit of carbon, then its abundance is about one half the solar
value. The weakness of the CO features explains why Shahbaz et al. (1999) were 
unable to clearly detect them in their low S/N spectrum. Thus, the suggestion
of Hynes et al. (2005) that the $K$-band luminosity of V616 Mon has a 
significant non-stellar component can be rejected. This is not to say, however,
that there is no contamination. The presence of He I and H I emission 
indicates continued accretion, as does the flatter than expected continuum.

\subsection{Determination of the $K$-band Contamination via Comparison with 
SS Cygni}

Both weak CO features and red continuua were common in our infrared 
spectroscopic survey of long period CVs (Harrison et al.  2004). {\it In fact, 
the slope of SS Cyg's $K$-band spectrum is very similar to that of V616 Mon}.
SS Cyg is a dwarf nova with an orbital period of 6.6 hr, and whose quiescent
accretion rate is about 4.8 $\times$ 10$^{\rm -11}$ M$_{\sun}$ yr$^{-1}$ 
(Wheatley et al.  2003). This should be compared with an estimated 
accretion rate for V616 Mon of $\leq$ 3.0 $\times$ 10$^{\rm -11}$ 
M$_{\sun}$ yr$^{-1}$ (McCintock et al. 1983). Along with nearly identical 
spectra, both secondary stars have similar masses: 0.70 M$_{\sun}$ for SS Cyg 
(Friend et al. 1990), and 0.68 M$_{\sun}$ for V616 Mon (Gelino et al. 2001), 
and line of sight rotation rates of 83 km s$^{\rm -1}$ for V616 Mon (Marsh et 
al. 
1994), and 87 km s$^{\rm -1}$ for SS Cyg (Martinez-Pais et al. 1994). As shown 
in Fig. 4, the $JHKL'$ light curves of SS Cyg exhibit ellipsoidal variations of 
similar amplitude to those observed for V616 Mon (see Fig. 3 in Gelino et al. 
2001). The orbital inclinations of the two binaries are also quite similar. 
Thus, one can make the case that the quiescent accretion rate in V616 Mon 
might not be too different from that for SS Cyg. The question is by 
how much does this accretion luminosity actually dilute the ellipsoidal 
variations of these two objects?

The key to estimating the contamination level requires observations at
longer wavelengths, where the flatter continuum component can become dominant
over the rapidly falling Rayleigh-Jeans tail of the secondary star. 
Unfortunately, such data do not yet exist for V616 Mon. If we continue with the 
SS Cyg analogy, an object that does have mid-infrared photometry, we can derive 
both the spectrum and contamination level of its accretion 
disk emission. In Fig. 5 is the $UBVRIJHKL'M$ spectral energy distribution 
(SED) of SS Cyg in quiescence from Dubus et al. (2004). Additionally, Dubus et 
al. have used Keck to obtain 11.7 $\mu$m observations of SS Cyg. It is clear 
that there is a mild mid-infrared excess above the SED of an isolated K4V 
secondary star. We find that the simplest model for the observed SED of SS Cyg 
is one in which a minor component due to free-free emission is added to the SED 
of the cool secondary star. If we use the ``non-flaring'' 11.7 $\mu$m  flux of 
SS Cyg, we find that the free-free component accounts for 4\% of the $K$-band
flux in SS Cyg. [If we include the flaring flux level, the free-free 
component is responsible for 13\% of the $K$-band luminosity. The origin
of this short time scale flaring has not been identified.] Thus, 
the ellipsoidal variations of SS Cyg are not significantly contaminated. 
Assuming an identical contamination level for V616 Mon, and using the
model parameters from Gelino et al. (2001), we find that a 4\% dilution
of the $K$-band light curve leads to an orbital 
inclination angle of 44$^{\circ}$. This implies a mass for the black hole of 
M$_{\rm 1}$ $\geq$ 9.2 $\pm$ 1.9 M$_{\sun}$. However, it is important to note 
that the $JHK$ light curves of SS Cyg reveal direct evidence for contamination 
with significant deviations from the light curve model, and asymmetric minima 
in both $J-$ and $H-$bands. Such features are not seen in the multi-epoch data 
presented by Gelino et al. (2001), but were present in the infrared light curves 
published by Froning \& Robinson (2001), suggesting some variability in the 
accretion rate for V616 Mon. 

\subsection{The Evolutionary History of the Secondary Star in V616 Mon}

Gonz$\acute{a}$lez Hern$\acute{a}$dez et al. (2004) have discussed the scenarios in which
the secondary star could have become polluted by the accretion of SN ejecta
assuming the currently observed secondary star mass is similar to its initial
mass (M$_{\rm 2{\rm initial}}$ $\leq$ 1.0 M$_{\sun}$). For their models where the
black hole mass is M$_{\rm 1}$ $>$ 11 M$_{\sun}$, they could achieve
the Al and Fe abundance enhancements through efficient capture of material
enriched by nucleosynthesis in the SN eruption. In those models, however,
they did not specifically analyze the carbon abundance, but assumed that it 
would track the enrichment of iron. They did note the possibility that O, Mg
and C-enhanced materials might be preferentially ejected in the equatorial
plane. Given the observed carbon deficit, the accretion of significant
SN ejecta seems less plausible.

In Harrison et al. (2005) we discussed the apparent need for a revision
in the evolutionary history of CVs due to our observations of carbon
deficits, and apparent deficits/enhancements of other species, such as Mg and
$^{\rm 13}$C. In the case of CVs, the secondary stars in the
pre-CV binaries were originally proposed to have always been of low mass, and relatively
unevolved. However, if higher initial masses are assumed, the observed abundance
 anomalies could have 
simply been created through normal evolutionary effects in the secondary star 
over the lifetime of the binary system. If the initial mass of the secondary
star was high enough (M$_{\rm 2_{\rm initial}}$ $\geq$ 1.3 M$_{\sun}$) to ignite the 
CNO cycle, an extended period of mass transfer could have
stripped the outer envelope of this object, revealing deeper layers 
where the CNO cycle once had been active. Certainly, such a model could also 
apply to V616 Mon.

The formation and evolution of black hole binaries has been re-examined
by Podsiadlowski et al. (2003). They confirm the results from earlier studies
 that it is quite difficult to
produce short period binaries containing a black hole and a low mass secondary
star. However,  Podsiadlowski et al. found that they could produce substantial
numbers of binaries with black hole primaries in models that started with 
intermediate mass secondary stars (2 $\leq$ M$_{\rm 2_{\rm initial}}$ $\leq$ 20 M$_{\sun}$). In addition,
for the highest mass secondary stars, they found that they could both grow the 
mass of the black hole, and produce a low mass X-ray binary with a chemically 
peculiar secondary star, {\it one that would show evidence for CNO processing}. 
In such models, dramatic increases in the mass of the black hole (of up 
to 7 M$_{\sun}$) were possible, even though they assumed Eddington-limited 
accretion. Given our evidence for CNO cycling in the secondary star of V616 
Mon, its enhanced
abundances of other metals, and the high mass primary, it seems likely that its 
secondary began life with a mass substantially larger than what is observed
today. 

It is critical to determine the level of the nitrogen and $^{\rm 13}$C 
enhancements
to fully explore the case for CNO cycle processed material in the secondary 
star of V616 Mon to derive limits on its initial mass. For
CVs, Harrison et al. (2005) noted the strong correspondence between extreme
N V/C IV line ratios seen in UV spectra, with the observed deficits of carbon
seen in IR spectra.
It is interesting to note that the black hole system XTE J1118+480 has
an extreme N V/C IV ratio, suggesting CNO processed material is being
 transfered to the  $\sim$ 7 M$_{\sun}$ black hole in that system (Haswell et 
al. 2002). Unfortunately, data of similar quality do not exist for V616 Mon, 
but our CV analogy suggests that it too should display an extreme N V/C IV 
ratio. In contrast, the black hole systems GU Mus (Shrader et al.
1993) and J0422+32 (Shrader et al. 1994) appear to have normal N V/C IV line 
ratios. Thus, we would expect that the secondaries in those systems would have 
normal CO features, and to be relatively unevolved. In this way they would 
be like the secondary 
stars in the magnetic CVs (``polars'', see Harrison et al. 2005). To explain 
the dichotomy seen in the secondary stars of magnetic and non-magnetic CVs, 
Harrison
et al. proposed that there must be two formation channels for those objects. 
It would be interesting if a similar result is found for close binaries 
containing black holes. Such a conjecture can be easily tested with further
UV and infrared spectroscopy of these systems.

\section{Conclusions}

We have obtained a new $K$-band spectrum of V616 Mon using NIRSPEC on
Keck II and find that the
secondary star is clearly detected, and appears to be a late-type K dwarf with
a significant $^{\rm 12}$C deficit. The weakness of the CO absorption 
explains the lack of detection of these features in an earlier spectrum, a 
result which had been used to argue for a significant contamination of the
system's 
$K$-band flux. While we do find that there is some contamination of the infrared 
spectrum in V616 Mon due to continuing accretion, it does not dramatically 
change previously estimated values 
for the orbital inclination, and hence the black hole mass for this system 
remains large. This source warrants further 
infrared spectroscopy to conclusively determine whether $^{\rm 13}$C is 
truly enhanced, and UV spectroscopy to investigate the N V/C IV ratio that is a 
strong indicator for the transfer of material that has undergone CNO cycle
processing.

\section{References}
\begin{flushleft}
Cannizzo, J. K. 1993, ApJ, 419, 318\\
Dubus, G., Campbell, R., Kern, B., Tamm, R. E., \& Spruit, AH. C. 2004, MNRAS, 349, 869\\
Froning, C. S., \& Robinson, E. L. 2001, AJ, 121, 2212\\
Gelino, D. M., Harrison, T. E., \& Orosz, J. A. 2001, AJ, 122. 2668\\
Gonz$\acute{a}$lez Hern$\acute{a}$ndez, J. I., Rebolo, R., Israelian, G., Casares, J., Maeder, A.
\& Meynet, G. 2004, ApJ, 609, 988\\
Goorvitch, D. 1994, ApJ Supp., 95, 535\\
Gray, R. O., \& Corbally, C. J. 1994, AJ, 107, 742\\
Harrison, T. E., Howell, S. B., Szkody, P., \& Cordova, F. A. 2005, ApJ, 632,
L123\\
Harrison, T. E., Howell, S. B., \& Osborne, H. L. 2004, AJ, 127, 3493\\
Haswell, C. A., Hynes, R. I., King, A. R., \& Schenker, K. 2002, MNRAS, 332,
928\\
Haswell, C. A., Robinson, E. L., Horne, K., Stiening, R. F., \& Abbott, T. M. C. 1993 ApJ 411, 802\\
Hynes, R. I., Robinson, E. L., \& Bitner, M. 2005, ApJ, 630, 405\\
Kallrath, J., Milone, E. E., Terrell, D., \& Young, A. T. 1998, ApJ, 508, 308\\
Kalogera, V. 1999, ApJ, 521, 723\\
Lee, C. -H., Brown, G. E., \& Wijers, R. A. M. J. 2002, ApJ, 575, 996\\
Leibowitz, E. M., Nemar, S., \& Orio, M. 1998, MNRAS, 300, 463\\
Marks, P. B., \& Sarna, M. J. 1998, MNRAS, 301, 699\\
Marsh, T. R., Robinson, E. L., \& Wood, J. H. 1994, MNRAS, 266, 137\\
Martinez-Pais, I. G., Giovannelli, F., Rossi, C., \& Gaudenzi, S. 1994, A\&A, 291, 455\\
McClintock, J. E., Petro, L. D., Remillard, R. A., \& Ricker, G. R. 1983, ApJ 266, L27\\
Menou, K. 2002, in ASP Conf. Proc. 262, The Physics of Cataclysmic Variables
and Related Objects, ed. B. T. G$\ddot{a}$nsicke, K. Beuermann, \& K. Reinsch (San Francisco: ASP), 387\\
Oke, J. B. 1977, ApJ, 217, 181\\
Podsiadlowski, Ph., Rappaort, S., \& Han, Z. 2003, MNRAS, 341, 385\\
Shahbaz, T., Naylor, T., \& Charles, P. A. 1994, MNRAS, 268, 756\\
Shahbaz, T., Bandyopadhyay, R. M., \& Charles, P. A. 1999, A\&A, 346, 82\\
Shrader, C. R., Wagner, R. M., Hjellming, R. M., Han, X. H., \& Starrfield, S. G. 1994, ApJ, 434, 698\\
Shrader, C. R., Gonz$acute{a}$lez-Riestra, Cheng, F. H., Horne, K., Panagia, N., Gilmozzi, R., \& Lund, N, 1993, A\&A Supp., 97, 309\\
Wallace, L., Livingston, W., Hinkle, K., \& Bernath, P. 1996, ApJ Supp., 106,
165\\
Wheatley, P. J., Mauche, C. W., \& Mattei, J. A. 2003, MNRAS, 345, 49\\
\end{flushleft}

\clearpage

\figcaption{The NIRSPEC data for V616 Mon (middle), compared with the IRTF + SPEX
data for two K dwarfs. Emission lines from He I at 2.06 $\mu$m,
and H I Br$\gamma$ at 2.16 $\mu$m are present in the spectrum of V616
Mon, but the strength of the H I feature is artificially enhanced due to
the telluric correction process. The template spectra have been convolved
to match the observed rotation rate of the secondary star in V616 Mon
($v$sin$i$ = 83 km $^{\rm -1}$). The strongest absorption features are
identified. Close comparison to the template spectra
suggests that the secondary star in V616 Mon has a spectral type of K5.}

\figcaption{A comparison of model spectra (black) generated using Kurucz
atmospheres and the spectral modeling program ``SPECTRUM'', compared to a
sequence of MK templates (red) covering the range K2V to M0V. The model spectra
have been smoothed to match the resolution of the SPEX instrument (5.1 \AA/pix).
}

\figcaption{A comparison of the spectrum of V616 Mon (red) to two model spectra
(black). The model spectrum on top has 50\% of the solar value of $^{\rm 12}$C,
and no $^{\rm 13}$C, while the bottom model has 40\% of the solar abundance
of $^{\rm 12}$C, and $^{\rm 13}$C/$^{\rm 12}$C = 1.0. We also identify the
location of a strong Fe I feature that can be clearly seen in the template star
spectra shown in Figs. 1 and 2.}

\figcaption{The $JHKL'$ light curves of SS Cyg obtained using SQIID on the KPNO
2.1 m, and NSFCAM on the IRTF. The $L'$ data were obtained at the IRTF
in 2004, while the $JHK$ data were obtained in 2003. Obviously, these data
were not simultaneous, but a $K$-band light curve of SS Cyg obtained at the same time as the $L'$ observations revealed the same amplitude ellipsoidal variations as seen in the SQIID $K$-band observations. The solid line is for a model with 
a K4V secondary star, and an orbital inclination angle of 40$^{\circ}$.}

\figcaption{The spectral energy distribution of SS Cyg (solid circles) from
Dubus et al. (2004) supplemented with our $L'$ photometry from the IRTF 
(+ NSFCAM). The error bars on the infrared data are plotted (the error bars
on $UBVRI$ data are $\leq$ 2\%, and are not plotted). The SED of a K4V, normalized so that it has 96\%
of the $K$-band flux of SS Cyg, is represented by the stars. Two blackbody
$+$ free-free models are plotted as solid lines. The lower one, fit to the
non-flaring 11.7 $\mu$m flux of SS Cyg, has 96\% of the $K$-band flux
originating in the blackbody component, while the ``flaring'' flux light
curve is normalized so that the blackbody has 87\% of the $K$-band flux.}

\begin{figure}
\plotone{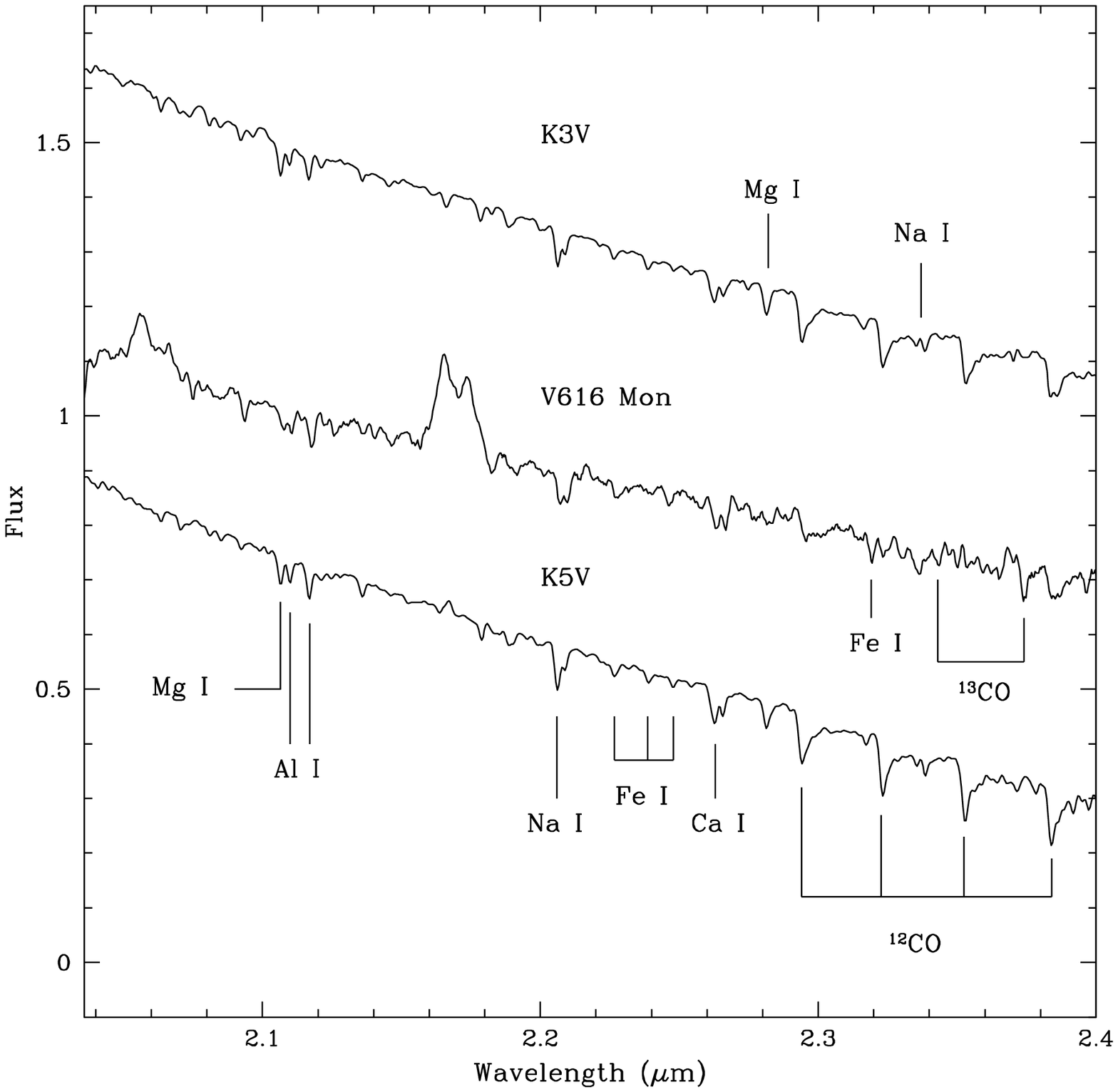}
\end{figure}

\begin{figure}
\plotone{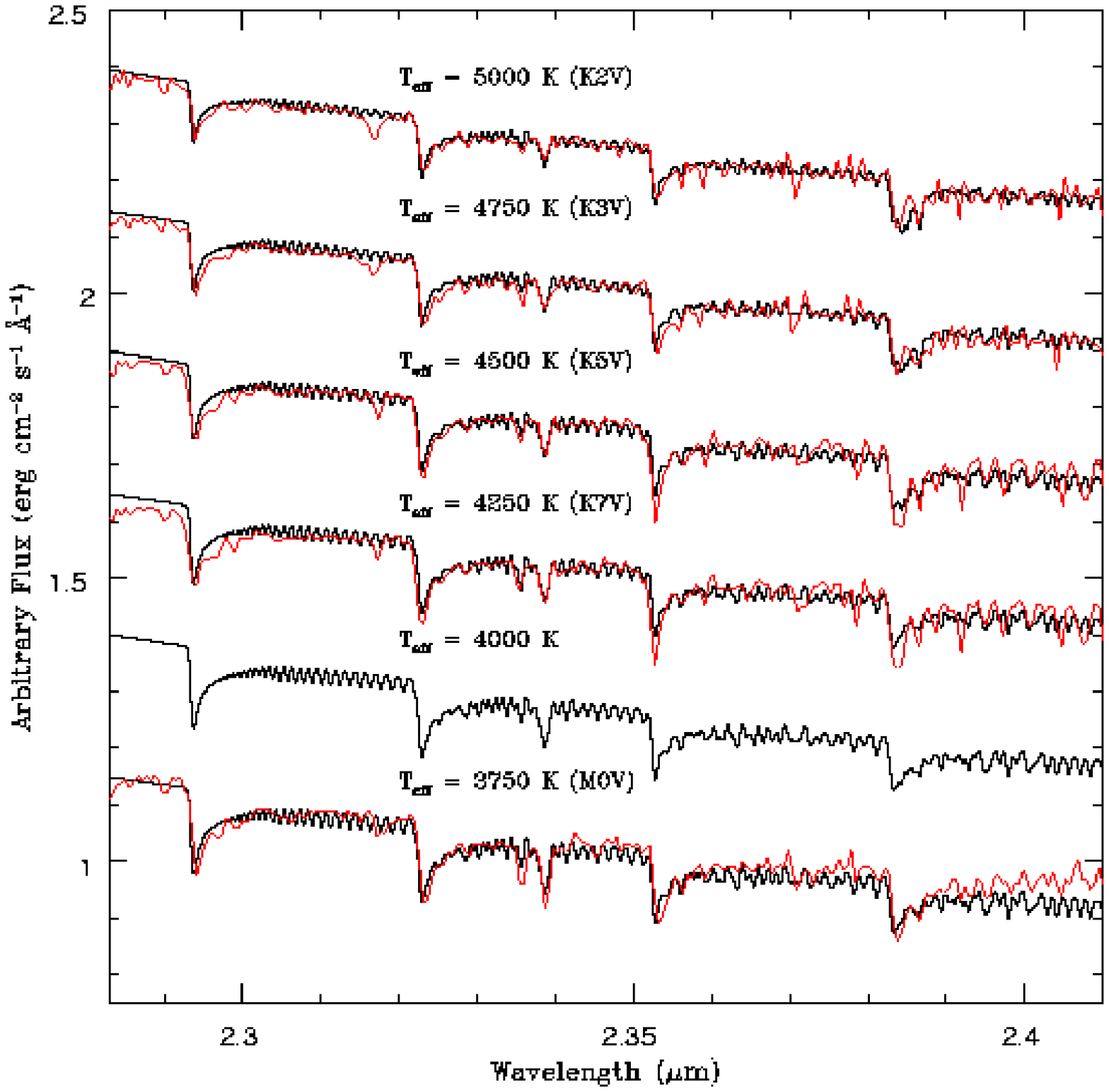}
\end{figure}

\begin{figure}
\plotone{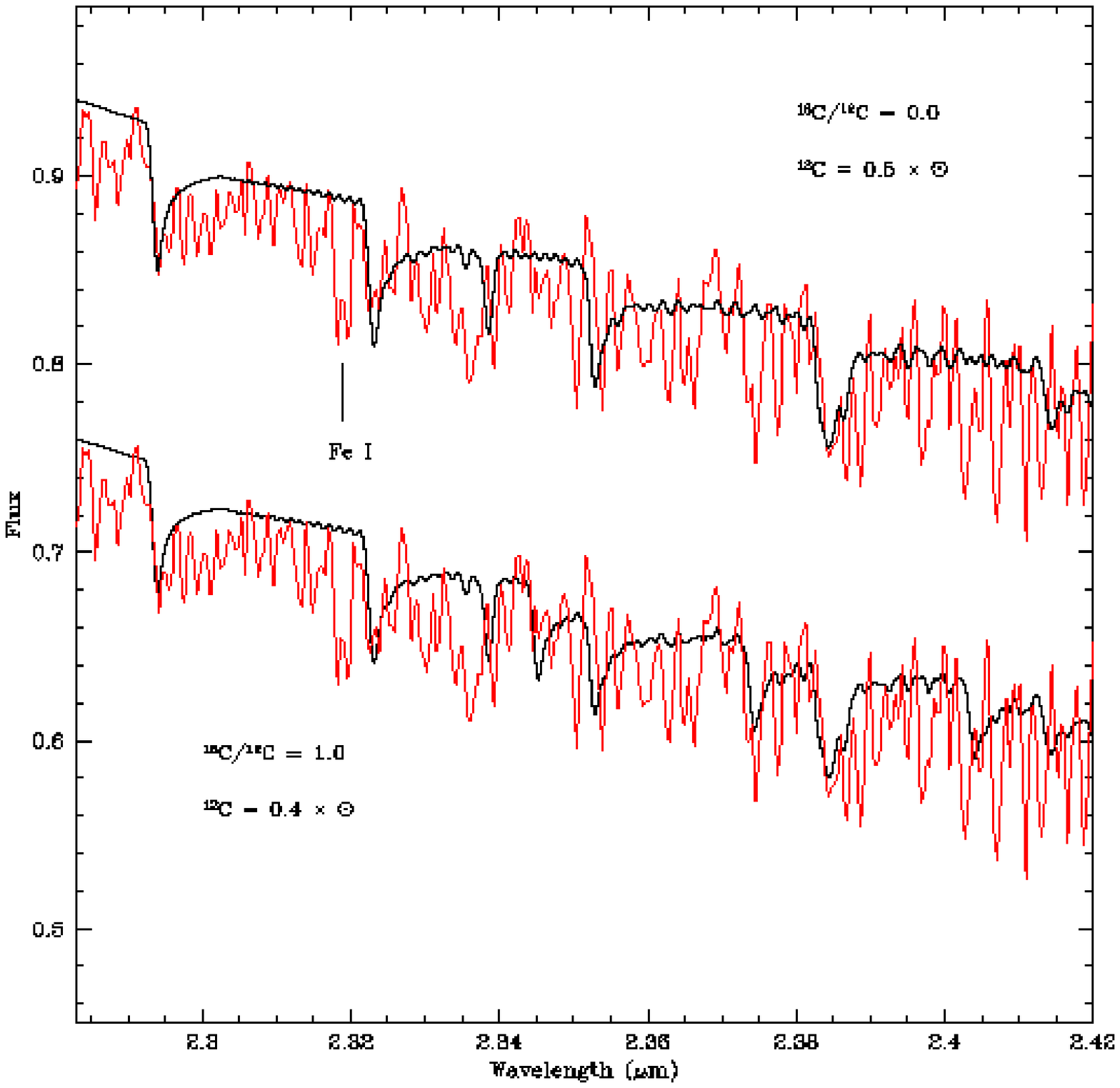}
\end{figure}

\begin{figure}
\plotone{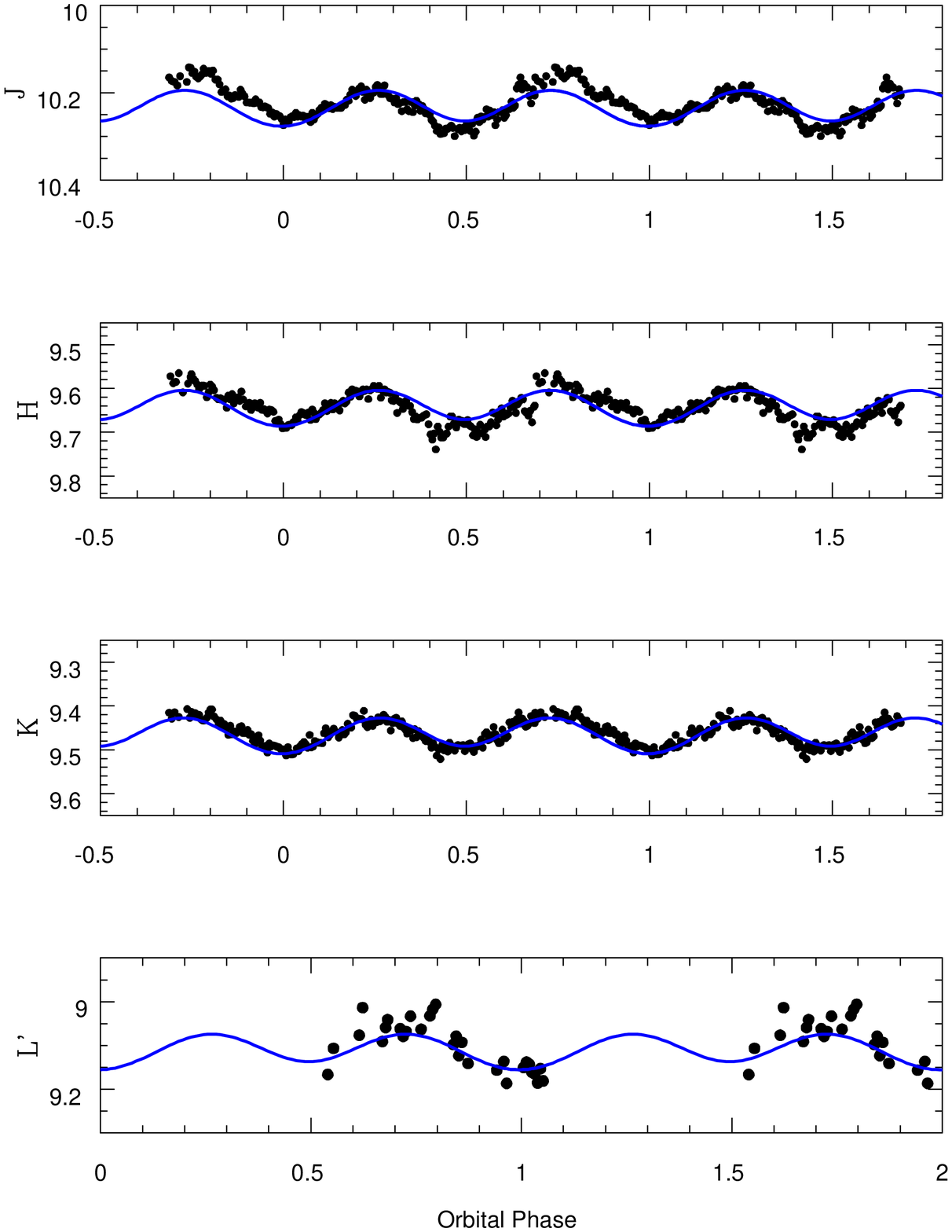}
\end{figure}

\begin{figure}
\plotone{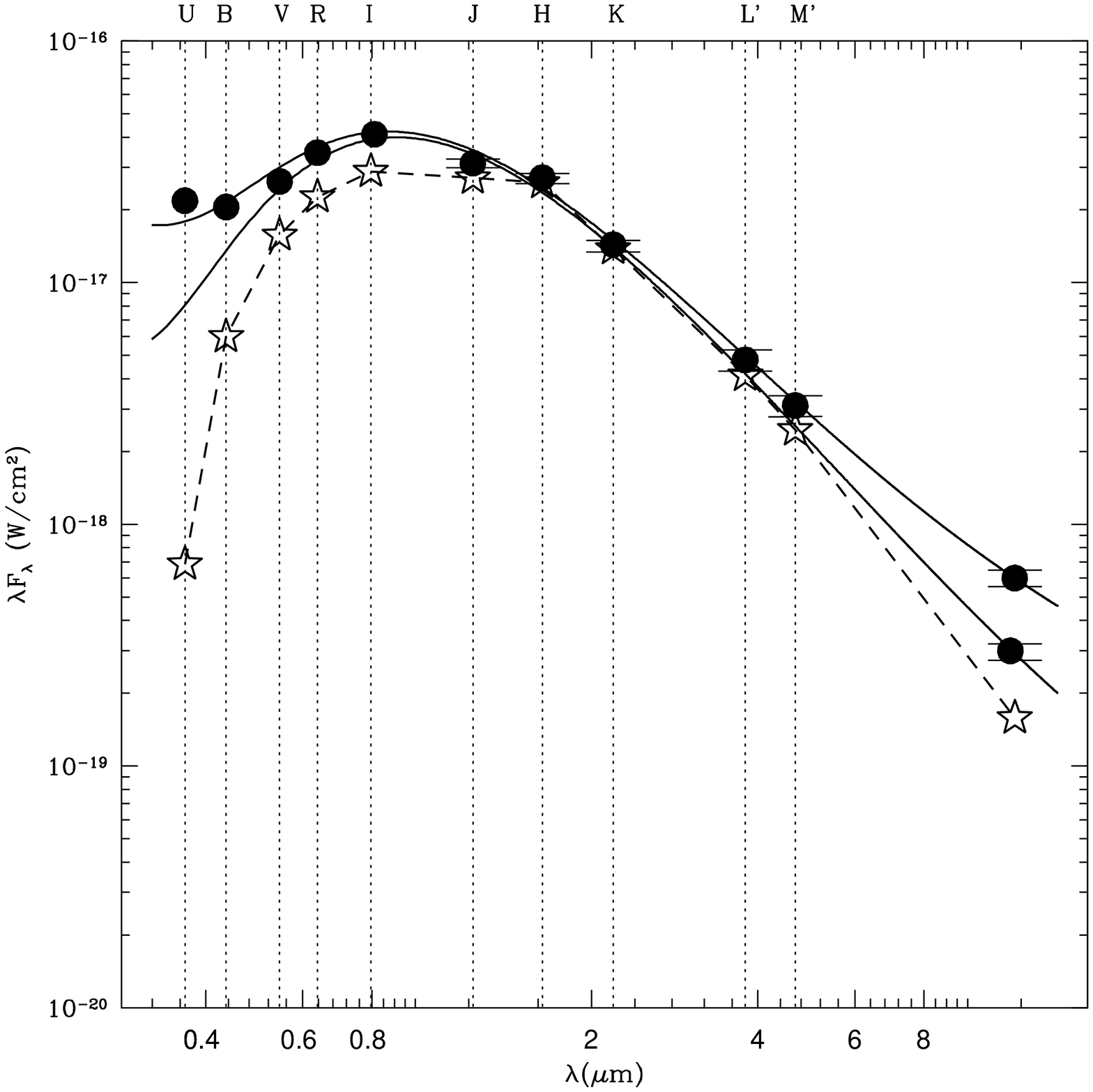}
\end{figure}

\begin{deluxetable}{ccccc}
\tablecolumns{5}
\tablewidth{0pc}
\tablecaption{Equivalent Widths}
\tablehead{\colhead{}&\colhead{}&\multicolumn{3}{c}{EqW (\AA)}\\
\colhead{Wavelength ($\mu$m)} & \colhead{Species} & \colhead{K3V} & \colhead{K5V} & \colhead{V616 Mon}}
\startdata
2.0735  &      Fe I      &  0.86$\pm$0.04 & 0.38$\pm$0.03 & 0.65$\pm$0.03 \\
2.0820  &  Si I + Fe I   &  0.69$\pm$0.03 & 0.57$\pm$0.03 & 1.46$\pm$0.07 \\
2.0927  &      Si I      &  0.49$\pm$0.04 & 0.36$\pm$0.02 & 0.76$\pm$0.04 \\
2.1065  &      Mg I      &  1.05$\pm$0.06 & 1.28$\pm$0.06 & 1.04$\pm$0.08 \\
2.1099  &  Al I + Fe I  &  0.59$\pm$0.04 & 0.63$\pm$0.04 & 0.94$\pm$0.07 \\
2.1169  &  Al I + Fe I  &  0.81$\pm$0.03 & 0.99$\pm$0.03 & 1.30$\pm$0.07 \\
2.1214  &      Al I      &  0.35$\pm$0.02 & 0.19$\pm$0.02 & 0.32$\pm$0.04 \\
2.1239  &     Fe I       &  0.07$\pm$0.02 & 0.19$\pm$0.02 & 0.34$\pm$0.05 \\
2.1360  &      Si I      &  0.35$\pm$0.03 & 0.50$\pm$0.04 & 0.56$\pm$0.03 \\
2.1460  & Fe I+Ca I+Mg I &  0.52$\pm$0.05 & 0.70$\pm$0.06 & 1.20$\pm$0.06 \\
2.1889  & Si I+Ti I+Fe I &  1.00$\pm$0.05 & 0.76$\pm$0.03 & 1.39$\pm$0.07 \\
2.2011  &      Ti I      &  1.09$\pm$0.05 & 0.41$\pm$0.04 & 0.30$\pm$0.06 \\
2.2076  &      Na I      &  2.49$\pm$0.07 & 3.39$\pm$0.09 & 3.32$\pm$0.09 \\
2.2264  &      Fe I      &  0.48$\pm$0.03 & 0.59$\pm$0.03 & 1.01$\pm$0.04 \\
2.2380  &      Fe I      &  0.58$\pm$0.03 & 0.47$\pm$0.03 & 0.62$\pm$0.05 \\
2.2642  &      Ca I      &  2.94$\pm$0.08 & 3.61$\pm$0.10 & 3.35$\pm$0.09 \\
2.2814  &      Mg I      &  1.54$\pm$0.05 & 1.55$\pm$0.05 & 1.45$\pm$0.08 \\
2.3165  &      Fe I      &  0.77$\pm$0.03 & 0.53$\pm$0.03 & 0.97$\pm$0.07 \\
\enddata
\end{deluxetable}
\end{document}